\documentstyle[12pt]{article}

\newcommand{\be}{\begin{equation}}
\newcommand{\ee}{\end{equation}}

\begin{document}

\begin{center}

{\Large {\bf Are Damage Spreading Transitions Generically in the
      Universality Class of Directed Percolation?}}

\vspace{1.3cm}

{\bf Peter Grassberger}

Physics Department, University of Wuppertal, D-42097 Wuppertal, FRG

\vspace{1.0cm}

\today

\vspace{1.3cm}

{\bf Abstract} \end{center}

{\small \advance \baselineskip by -2pt

We present numerical evidence for the fact that the damage
spreading transition in the Domany-Kinzel automaton found by
Martins {\it et al.} is in the same universality class as
directed percolation. We conjecture that also other damage
spreading transitions should be in this universality class,
unless they coincide with other transitions (as in the Ising
model with Glauber dynamics) and provided the probability
for a locally damaged state to become healed is not zero.
}

\vspace{2.cm}


\eject

Among all critical phenomena, directed percolation (DP) is maybe that
which has been associated with the most wide variety of phenomena.

First there are interpretations where the preferred direction is a
spatial direction. This was of course proposed to apply to material
and charge transport in disordered media under the influences of
external forces. Also, it should model the propagation of epidemics
and forest fires under some directional bias, e.g. strong wind.

More interesting are interpretations where the preferred direction
is time. Here, the primary interpretation is as an epidemic without
immunization, the so called ``contact process" \cite{liggett} or
the ``simple epidemic" \cite{mollison}.

But these are by no means all possible applications. A very early
application (even if it took rather long until it was understood
as such \cite{sund,cardy}) was to ``reggeon field theory", a
theory for ultrarelativistic particle collisions popular in
the 70's \cite{moshe}. Here, the preferred direction is that of
``rapidity", while the directions transverse to it are provided
by the impact parameter plane. This connection is interesting
since it was through it that first precise estimates of critical
exponents and amplitudes were obtained for DP \cite{moshe}.

Another realization of the DP transition occurs in simple
models of heterogeneous catalysis. The first such model was
proposed by Ziff {\it et al.} \cite{ZGB} (ZGB). The simulations
of these and subsequent authors indicated that this model was
in a different universality class, and it is only after some
controversies that it is now generally accepted to be in the
DP universality class \cite{jensen1}. Similar models are invented
again and again \cite{droz,albano}. Repeatedly they are claimed
to be in different universality classes, and repeatedly these
claims are refuted \cite{jensen,redner,unpub}.

In \cite{janssen,grass-schl} it was proposed that the universality
class of DP contains all continuous transitions from a ``dead"
or ``absorbing" state to an ``active" one with a single scalar
order parameter, provided the dead state is not degenerate
(and provided some technical points are fulfilled: short range
interactions both in space and time, nonvanishing probability
for any active state to die locally, translational
invariance [absence of `frozen' randomness], and absence of
multicritical points). It seems fair to say that there is now
ample evidence for this proposal. It predicts, e.g. immediately
that the ZGB model is in this universality class. A
rather subtle question is whether also chaotic systems where
the random noise is replaced by deterministic chaos are in the
same class \cite{pomeau,grass-schr}.

As far as I am aware of, no model with non-fluctuating absorbing
state and a multi-component order parameter has ever been studied
in the literature. Notice that this has to be distinguished
from models for which some {\it mean field approximation}
has a multi-component order parameter. Such models are quite
common (e.g. the Bethe-Peierls approximation of the Ising
model, or the mean field approximation of ZGB), and it was
just the study of such a model which had led to the
conjecture in \cite{grass-schl}. A supposed generalization
\cite{grin} of the above conjecture is thus already fully
contained in the original conjecture of \cite{grass-schl}.

A more interesting question is what happens if the dead state
is degenerate. Counter examples with twofold degeneracy were
studied in \cite{gkv,takayasu,jens6}. They involve conservation
laws which prevent some active states from dying, making it
thus immediately clear that any transition --- if it occurs at
all --- has to be in a different universality class.

But the main open problem is whether models can be generically
in the DP class if they have an absorbing state with positive
entropy (for obvious reason, we prefer not to call it ``dead"
in this case). One might conjecture that such a state is
essentially unique on a coarse scale, provided its evolution
is ergodic and mixing --- and provided it does not involve
long range correlations (long correlations should be entirely
due to patches of ``active" states). Since only coarse-grained
properties should influence critical behavior, this would
suggest that such transitions are in the DP class. This seems
contradicted by simulations of some catalysis models
\cite{avraham,yaldram}. But as we pointed out already, systematic
errors are often underestimated, and recent simulations of these
and similar models support the conjecture
\cite{jens2,jens-dick3,mendes,jens5} (as explained below, we
believe that violations of universality observed in some of the
latter papers for `dynamic' properties should be disregarded).

In this note we propose that there is a rather large and well
studied class of transitions which are exactly of the latter
type, and which are thus all in the DP class. These are so
called ``damaging" transitions. In these models one considers
two replicas of a stochastic spin system, and lets them evolve
with {\it identical realizations} of the stochastic noise.
The initial conditions can be either completely independent,
or one can start with two states which are identical except for
a single spin. This single flip is considered as a ``damage",
and the question is whether this damage will finally heal, so
that both replicas converge towards identical states --- or
whether it will spread. If the two states are uncorrelated
initially, the transition is between a situation where their
rescaled Hamming distance (= density of damaged sites) stays
finite and one where this distance goes to zero.

More precisely, we propose that such damaging transitions are
in the DP class if they do not coincide with another transition
(since then there would be long range correlations in the
absorbing state), and if there is no frozen randomness. The
former applies to the 2-d Ising model with Glauber dynamics,
since there the damaging transition coincides with the ordinary
critical point \cite{stanley} (the situation is less clear in
3 dimensions \cite{costa,caer}). Frozen randomness is involved
in damaging in spin glasses \cite{arcang,campbell,jan} and
in the extensive studies of damaging in Kauffman models
\cite{kaufmann,derrida-stauffer,da_silva}. This should be in
the same universality class as DP with frozen randomness
\cite{noest}, but for the Kauffman models there is a further
complication: there a damage typically does not heal completely,
whence the `dead' state is not absorbing in our sense. We might
mention that it was already pointed out that damaging in the
{\it annealed} Kaufmann model is in the DP class
\cite{derrida-stauffer}, but this is much more trivial than our
present claim. The annealed model can be mapped {\it exactly}
onto DP, which is not the case in general.

We support our claim with simulations of damaging in the
Domany-Kinzel cellular automaton (CA) \cite{domany}. This is
a CA with one space and one time dimension, and with two
states per site: $s_i = 0,1$. Dynamics is defined by the
following rule involving two real parameters
$p_1$ and $p_2$ (we make a trivial modification which
slightly simplifies the simulation): \\
\hspace{1.cm} (i) if $s_i=0$ and $s_{i+1}=0$ then $s_i'=0$ \\
\hspace{1.cm} (ii) if $s_i$ XOR $s_{i+1} =1$ then $s_i'=1$ with
    probability $p_1$ and $s_i'=0$ with probability $1-p_1$ \\
\hspace{1.cm} (iii) if $s_i$ AND $s_{i+1} =1$ then $s_i'=1$ with
    probability $p_2$ and $s_i'=0$ with probability $1-p_2$. \\
For $p_1<1/2$ and any $p_2<1$, it is obvious that any state
will converge towards the dead state $\ldots 000\ldots$.
Actually, this state is an attractor for all values of $p_1$
below a critical curve $p_1^c(p_2)$. This curve is indicated
as curve ${\cal C}$ in fig.1. To the right of ${\cal C}$,
one has an active state (the dead state still is stationary,
but it no longer attracts all initial states) with
$\rho\equiv \langle s_i\rangle>0$.

The above conjecture suggests that the transition all along
${\cal C}$ is in the DP class, except at its upper limit
point $(p_1,p_2)=(1/2,1)$ where the model is a discrete time
variant of the exactly solvable voter model \cite{liggett}
(``compact directed percolation" \cite{essam1}). This
is supported by all numerical evidence \cite{kinzel,kohring}
(except for a renormalization group analysis and Monte Carlo
simulations presented in \cite{livi}; in high precision
Monte Carlo simulations \cite{unpub} we could not confirm
these claims). In particular, bond and site DP correspond
to $p_2=(2-p_1)p_1$ and $p_1=p_2$, respectively.

It was found recently in \cite{martins} that the active
phase can be further subdivided into a phase in which damage
does not spread (``healing active phase") and one where
it does (``chaotic"). The transition between these two
phases is indicated by curve ${\cal D}$ in fig.1. It
corresponds to $p_1=p_1^d(p_2)$ where $p_1^d>p_1^c$
for all $p_2>0$, while $p_1^d(0)=p_1^c(0)$
\cite{rieger}. Indeed, one can consider two different
variants of the damaging process: in the first one uses
different random numbers when applying rules (ii) and
(iii) above, in the second one uses the same. Curve
${\cal D}$ is computed with the second variant. The first
variant would give a different curve slightly to the left
of ${\cal D}$ \footnote{The very existence of these two
variants shows that it is misleading to speak of different
phases in the Domany-Kinzel CA, as done in \cite{martins}.
Instead these are different phases for very specific
algorithms for simulating pairs of such automata.}.

As pointed out in \cite{kohring},
one can describe a pair of replicas by an extended
phase space with 4 states per site: (00), (01), (10) and
(11). Damage spreading corresponds then to the (directed)
percolation of states (10) and (01), while any state
with (00) and (11) only is healed. Since
$p_1^d>p_1^c$ for all $p_2>0$, the healing state has
positive entropy at the damage spreading transition,
and it does not immediately follow from
the conjecture of \cite{janssen,grass-schl} that this
transition is in the DP universality class.

To check our conjecture that it is in this class
nevertheless, we performed extensive simulations at
$p_1=1$ where both variants coincide. Less extensive
runs were made at several other values of $p_1$,
where we studied both variants.

We worked on lattices of length $L$ with periodic
boundary conditions.
To speed up our simulations, we simulated 64 lattices
simultaneously (we worked on machines with 64 bit
words) by assigning the $k$-th bit of the $i$-th
word in an integer array of length $L$ to the spin
$s_i^{(k)}$ in the $k$-th lattice. The dynamics is then
easily implemented by standard bit operations.

To measure the degree of damage in simulations which
start with independent random initial configurations
(thus with half of the sites damaged initially), we count
the number of set bits in each word. If this number
is $n_i$ for the $i$-th word, then the number of pairs of
lattices which are damaged at site $i$ is $(64-n_i)n_i$.
The sum of Hamming distances between all $64\times
63/2=2016$ pairs of lattices is thus
\be
   d = \sum_{i=1}^L \;(64-n_i)n_i \;.
\ee
For simulations with initial single site damage, this
is not possible since we cannot build an initial state
in which {\it each} pair is damaged at only one site.
Instead, we introduced single site damage only between
successive bits, i.e. we initially damaged the
$(2k+1)$-st bit ($k=0,1,2,\ldots 31$) in 32 different
words, and counted how often the $(2k+1)$-st bit
differed from the $(2k)$-th one.

Results from runs with random and independent initial
states on lattices of size $L=2^{22}$ are
presented in fig.2. There we show the total damage
as function of time, for $p_1=1$ and several values of
$p_2$. At the critical point we expect an algebraic
decay, corresponding to a straight line in fig.2.
If the transition is in the DP class, this decay
is governed by an exponent $\delta=0.1596\pm 0.0001$
\cite{baxt-gut,essam,dick-jens}. We see indeed a nearly
perfect straight line for $p_2\approx 0.3122$. Together
with the data described below this gives our estimate
\be
   p_2^d=0.31215\pm 0.00004 \qquad {\rm for} \quad p_1=1\;,
\ee
and the exponent extracted from it ($0.157\pm 0.002$)
is in good agreement with DP. Similar results (although
somewhat less precise) were obtained for both variants of
the damage spreading at several values of $p_1$. For
$p_1=0.85$, e.g., they gave $p_2^d=0.1957\pm 0.0002$
(variant 1) resp. $p_2^d=0.1400\pm 0.0002$ (variant 2).
In all cases $\delta$ was compatible with the DP value.

It is well known from studies of DP that the exponent
$\beta$ defined by
\be
   d \sim (p_2^d-p_2)^\beta
\ee
is not easily measured precisely due to the very long
transients close to the critical point (i.e., due to
the smallness of $\delta$) and due to finite size
effects. The latter are absent in our simulations due
to the very large lattice size. Nevertheless,
extrapolating the data from fig.2 to $t\to\infty$ gave
only a crude estimate $\beta=0.272\pm 0.006$ which
is however in perfect agreement with DP where $\beta
= 0.2766\pm 0.0003$ \cite{baxt-gut,essam,dick-jens}.

In order to measure an independent critical
exponent, we made in addition runs with initial single
site damage on much smaller lattices ($L\leq 7000$)
and for shorter times ($t\leq 40000$). Apart from the
damaged sites, the initial configurations were randomly
chosen active states (they were set to the final
configuration of the first lattice in the preceding
run by setting the $i$-th word to 0 if $s_i^{(1)}=0$,
and to -1 if $s_i^{(1)}=1$). From universality
with DP we expect that at the critical point
\be
   d \sim t^\eta, \qquad \eta=0.314\pm 0.001\;,
                         \label{spread}
\ee
which is nicely fulfilled. Off the critical point we
should have
\be
   \langle d\rangle \propto t^{-\delta}
      \phi((p_2^d-p_2)t^{1/\nu_{||}})
      \qquad\mbox{(full initial damage)}
                       \label{spread1}
\ee
and
\be
   \langle d\rangle \propto t^\eta
      \psi((p_2^d-p_2)t^{1/\nu_{||}})
      \qquad\mbox{(single site initial damage)}
                       \label{spread2}
\ee
with universal scaling functions $\phi(z)$ and $\psi(z)$
which are
regular at $z=0$, and with $\nu_{||}=1.7336\pm0.0005$.
To see that our data are fully consistent with this, we
plotted in fig.4 $d/t^\eta$ against $(p_2^d-p_2)
t^{1/\nu_{||}}$ for both types of initial conditions. We
see indeed a perfect data collapse
as predicted by the above ansatz. We just mention that
similar results (again with somewhat smaller statistics
and with significantly larger corrections to scaling)
were obtained for several other values of $p_1$, and
allowed us to locate curve ${\cal D}$ in fig.1 with high
precision.

In conclusion we have given numerical evidence that the
damage spreading transition in the Domany-Kinzel CA is
in the DP universality class, although the
undamaged state has positive entropy. We expect this
to be true in general, not only for the Domany-Kinzel CA.

Of course, we have to set initial conditions such that we
are not confined to atypical states carrying zero measure.
In the present case, such atypical behavior would e.g.
result if we would start with one of the configurations
being dead (all $s_i=0$) or nearly dead ($s_i\ne 0$ only
in a finite region). In the latter case, we would then
have a linear increase of $d$ instead of eq.(\ref{spread}).
We believe that not taking into account this caveat is the
reason why only partial universality with DP was observed
in \cite{jens-dick3,mendes}. In these papers, `dynamical'
simulations were done where the active region was bounded
and expanding with time. Outside this region the configurations
were not allowed to evolve, but were (artificially) kept
in atypical states. It seems trivial that this modification
of the model can lead to violations of universality.

Unfortunately, our conjecture does not immediately apply to
the case of Kauffman automata \cite{kaufmann} where damage
spreading had been studied quite intensively
\cite{derrida-stauffer,da_silva}. First of all, these
models involve frozen randomness and should thus --- if at
all --- be compared to DP in disordered media. Secondly,
healing is not perfect in Kauffman models even in the phase
in which damage does not spread. In this phase a finite
damage has a non-zero probability to persist forever, and
the healed state is not absorbing in our sense. It would be
most interesting to study modified Kauffman models where
such healing takes place (e.g. stochastic versions ---
apart from the randomness in the attribution of local rules,
Kauffman models are strictly deterministic), and to compare
them with DP in disordered media.

We have added one more item to the already long list of
possible physical realizations of the DP transition. It is
vexing that in spite of this ubiquity in {\it models}, and in
spite of its conceptual simplicity (DP is by far the simplest
critical phenomenon to study on a computer and to explain to
a high school student), there have not yet been reported any
experiments where the critical behavior of DP was observed
even crudely! Maybe the present realization can lead to such
an observation.

On the more practical side, we have introduced a new and very
efficient method for simulating damage spreading which might
find applications also in similar problems.

Acknowledgement: I am most indebted to Dr. M. Schreckenberg
for interesting discussions and for pointing out the two
variants of implementing damage spreading in the
Domany-Kinzel CA, and to Dr. T. Lippert for help with running
the program on a Thinking Machines CM5. This work was
supported by DFG, SFB 237.

\eject

\vspace{1.3cm}

\section*{Figure Captions:}

{\bf Fig.1:} Part of the phase diagram for the Domany-Kinzel
CA. Curve ${\cal C}$ separates dead (left) from active
(right) phases. Curve ${\cal D}$ (which joins ${\cal C}$
at its lower end point, but runs otherwise entirely in
the active phase) separates a healing phase (left) from a
chaotic phase (right) where any damage has non-zero
probability not to heal. Here the damaging is implemented
according to the first variant described in the text. With
the second variant, ${\cal D}$ would be somewhat further
to the left. The transition curves were determined by runs
with single active/damaged initial sites, and demanding
that the exponent $\eta$ (see eq.(\ref{spread})) has the
value of DP. The precision of the curves is everywhere better
than the thickness of the lines. Quantitatively, the phase
diagram agrees with data from \cite{rieger}, but not with
the diagram given in \cite{martins}. It also deviates
significantly from that in \cite{zebende}.

{\bf Fig.2:} Log-log plot of the total number of damaged sites
in 2016 pairs of lattices, with $2^{22}$ sites each. For all
curves $p_1=1$, while $p_2$ ranges from 0.3086 to 0.3155
(from top to bottom). Initial configurations were random.

{\bf Fig.3:} Log-log plot of the total number of damaged
sites in runs where each pair of lattices was initially
damaged at a single site. Again $p_1=1$. The four central
values  of $p_2$ are the same as in fig.2.

{\bf Fig.4:} The same data as in figs.2 (panel a) and 3
(panel b), but plotted such that all data should collapse
onto a single curves if eqs.(\ref{spread1},\ref{spread2})
are correct. Only data for $t>40$ are plotted in panel a,
and for $t>10$ in panel b, in order to reduced finite-time
corrections.

\eject

\end{document}